\documentclass[reqno]{amsart}
\usepackage{amssymb,amsmath,amsthm}

\newtheorem{rem}{Remark}[section]
\newcommand{\br}{\begin{rem}}
\newcommand{\er}{\end{rem}}
\newtheorem{ex}{Example}[section]
\newcommand{\bex}{\begin{ex}}
\newcommand{\eex}{\end{ex}}
\newtheorem{Def}{Definition}[section]
\newcommand{\bd}{\begin{Def}}
\newcommand{\ed}{\end{Def}}
\newtheorem{theorem}{Theorem}[section]
\newcommand{\bt}{\begin{theorem}}
\newcommand{\et}{\end{theorem}}
\newtheorem{corollary}{Corollary}[section]
\newcommand{\bc}{\begin{corollary}}
\newcommand{\ec}{\end{corollary}}
\newtheorem{lemma}{Lemma}[section]
\newcommand{\bl}{\begin{lemma}}
\newcommand{\el}{\end{lemma}}
\newcommand{\be}{\begin{equation}}
\newcommand{\ee}{\end{equation}}
\newcommand{\bea}{\begin{eqnarray}}
\newcommand{\eea}{\end{eqnarray}}

\newtheorem{prop}{Proposition}[section]
\newcommand{\bpr}{\begin{prop}}
\newcommand{\epr}{\end{prop}}
\newcommand{\ds}{\displaystyle}
\newcommand{\bpf}{\begin{proof}}
\newcommand{\epf}{\end{proof}}

\title{On non-QRT Mappings of the Plane}
\author{P. Kassotakis}
\address{School of Mathematics and Statistics F07,
The University of Sydney, NSW 2006, Australia}
\email{pavlos@maths.usyd.edu.au, pavlos1978@gmail.com}
\author{N. Joshi}
\address{School of Mathematics and Statistics F07,
The University of Sydney, NSW 2006, Australia}
\email{nalini@maths.usyd.edu.au}
\date{\today}
\begin{document}

\maketitle

\begin{abstract}
We construct 9-parameter and 13-parameter dynamical systems of the plane which map bi-quadratic curves to other bi-quadratic curves and return to the original curve after two iterations. These generalize the QRT maps which map each such curve to itself. The new families of maps include those that were found as reductions of integrable lattices by \cite{Nalini1hky}.
\end{abstract}

\section{Introduction}

The QRT maps were proposed by Quispel, Roberts and Thompson \cite{QRT3} about twenty years ago and provided a fruitful starting point for the field of discrete integrable systems. In particular, the de-autonomization of these maps through the singularity confinement method led to the discrete Painlev\'e equations, which turned out to have very rich connections with other fields, including random matrix theory.

However, although the QRT mappings are considered as the most general family of Liouville integrable bi-rational maps of the plane, there exist examples of integrable maps that do not fall in this class. We consider a generalization of QRT maps, which have been introduced under the name of HKY mappings. Their non-QRT nature arises from the fact that they preserve a bi-quartic rational integral instead of a bi-quadratic one. Actually, the iterates of these systems alternate between $2$ birationally equivalent elliptic curves, hence their non-QRT nature. Several  examples of such  systems were exhibited  in \cite{Hirota1hky} by Hirota,  Kimura and  Yahagi, where they presented third-order mappings, which can be integrated to second-order with bi-quartic invariants, hence the initials HKY under which these examples are known. Nevertheless, the first researchers that gave examples of non-QRT type mappings were  Haggar, Byrnes, Quispel and Capel in \cite{quispel-hky}. For this reason, in this paper,   we describe these systems as non-QRT mappings.  More examples of non-QRT systems were generated by autonomisations of discrete Painlev\'e equations \cite{Hirota2hky},  reductions on integrable lattices \cite{Nalini1hky},  using an elliptic function solution \cite{viallet1hky}, solutions of Adler's lattice equation \cite{james-hky}, or by a direct approach \cite{gram_ram_11,tsuda_2,al-pal}.

In this paper, we extend these known examples by providing multi-parameter families of non-QRT mappings. Our main idea is to construct birational mappings $\phi$ which transforms an integral $I(x,y)$ of a QRT mapping to an involutive homography of itself. The construction then leads to mappings that preserve a bi-quartic integral and hence lead to non-QRT mappings. This construction is explained in further detail in Section 2. In Sections 3 and 4, multiparameter families of examples of non-QRT Liouville integrable mappings of the plane are presented. Finally, in Section 5, we discuss the integration of these systems, followed by conclusions in Section 6.

\section{From QRT to non-QRT Mappings}
The QRT mapping is defined through a biquadratic invariant
\[
{\ds I(x,y)=\frac{{\bf X}^TA_0{\bf
Y}}{{\bf X}^TA_1{\bf Y}}},
\]
where ${\bf X}$, ${\bf Y}$ are vectors ${\bf X}=(x^2,x,1)^T,$ ${\bf Y}=(y^2,y,1)^T$ and $A_0,$ $A_1$ are two  $3\times 3$ matrices,
\[
{\ds A_i=\left(\begin{array}{ccc}
\alpha_i& \beta_i &\gamma_i\\
\delta_i& \epsilon_i& \zeta_i\\
\kappa_i& \lambda_i & \mu_i
\end{array}\right)}.
\]
If these matrices are symmetric, i.e., $A_i=A_i^T$, the QRT mapping is called symmetric since its invariant is symmetric under the interchange of $x$ and $y$. If they are antisymmetric, i.e., $A_i=-A_i^T$, the QRT is called antisymmetric, otherwise it is called asymmetric. In the antisymmetric and asymmetric cases, the QRT is the composition $i_2\circ i_i$ of the non-commuting involutions $i_1,$ $i_2,$ where the latter are defined by the solution of the equations $I({\tilde x},y)-I(x,y)=0$ and $I(x,{\overline y})-I(x,y)=0$ respectively. So we have
\be \label{QRT}
i_1:\left\{ \begin{array}{l}
{\tilde x}={\ds \frac{f_1(y)-f_2(y)x}{f_2(y)-f_3(y)x}}\\[3mm]
    {\tilde y}=y
    \end{array}\right.,\quad
i_2:\left\{\begin{array}{l}
{\overline x}=x\\
    {\overline y}={\ds\frac{g_1(x)-g_2(x)y}{g_2(x)-g_3(x)y}}
    \end{array}\right.,\ee where  $$
\begin{array}{c}
{\ds (f_1(y),f_2(y),f_3(y))^T=(A_0{\bf Y})\times(A_1{\bf Y})},\\[3mm]
{\ds  (g_1(x),g_2(x),g_3(x))^T=(A_0^T{\bf X})\times(A_1^T{\bf X})}.
    \end{array}
$$

In the symmetric case, the QRT is defined by the composition of $i_1$ or $i_2$ with the involution $j:$
${\overline x}=y,$ ${\overline y}=x,$ that clearly preserves $I(x,y)$ since the latter is preserved under the interchange of its arguments. So the symmetric QRT can be written as the $3-$point map:
$${\ds {\overline y}=\frac{f_1(y)-f_2(y){\underline y}}{f_2(y)-f_3(y){\underline y}}}.$$

The results of \cite{Nalini1hky} suggest a direct approach to finding non-QRT mappings from the QRT integral. Specifically, it was shown that if one can find the appropriate $I(x,y)$ or equivalently the appropriate $A_i$ such that a birational mapping $\phi$ transforms  $I(x,y)$ to an involutive homography of itself, namely if
\begin{equation}\label{homography} {\ds I\circ \phi=\frac{aI-b}{cI-a}},\end{equation}
then $\phi $ will preserve the bi-quartic ${\ds I+\frac{aI-b}{cI-a}}$ and will be an non-QRT map. Note that in the case when $\phi$ is an involution,  or more generally a periodic map, one can get nontrivial non-QRT systems by composing $\phi$ with the QRT involution $i_1$ or $i_2$. Note that in this case, the non-commutativity of $i_1$ or $i_2$ with $\phi$ should be checked.

The authors of \cite{Nalini1hky} identified two cases, which we call {\em Type I}  and {\em Type II} non-QRT maps, where
\[
\begin{split}
{\rm Type\ I}:&\ I \circ \phi = -I\\
{\rm Type\ II}:&\ I \circ \phi = 1/I
\end{split}
\]
Equation (\ref{homography}) can be mapped to each of these two types.  If $c=0$, type I can be found after substituting $I\rightarrow I - b/a$. If  $c\not=0$, we get type II
after substituting $I\rightarrow I \sqrt{a^2/c^2+b/c} + a/c$. Therefore, the conserved quantities of $\phi$ are
$I^2$ and $I+1/I$, respectively.  These two families of mappings at a first glance appear to be different,  but are related \cite{james-com}, since if $\phi $ gives a {\em Type I} invariant, after defining $K={\ds \frac{1-I}{1+I}}$ , we get $K \circ \phi = 1/K.$  So, from now on, we deal only with {\em Type I} non-QRT mappings.

Note also that  for both types of non-QRT maps, their even (odd) iteration is a QRT mapping, since after two consecutive actions of the map, the bi-quadratic is mapped to itself, $I\rightarrow -I \rightarrow I$ or $I\rightarrow 1/I \rightarrow I.$ In this sense non-QRT maps can be described as a \lq\lq square root\rq\rq\ of the associated QRT map.

\section{Two Families of Antisymmetric non-QRT Maps}
Many examples of $3$-point non-QRT maps $(\phi_i)$ that appear in the literature are defined for symmetric $I(x,y)$.
That is, $\phi_i=j\circ h_i,$ where
$$
h_i:\left\{ \begin{array}{l}
  {\tilde x}=F_i(x,y)\\
 {\tilde y}= y
\end{array} \right., \quad
j:\left\{ \begin{array}{l}
  {\overline x}=y\\
 {\overline y}= x
\end{array} \right.,\quad \mbox{and} \quad h_i^2=j^2=id
$$ with $j$ of QRT type and $h_i$ of non-QRT type (without loss of generality of type I).

A natural question then arises: for which bi-quadratic  $I(x,y)$ is the involution $j$ of non-QRT type? If we can find such an $I$ then by composing $j$ with a QRT involution $i_1$ or $i_2$ of section (2), then we will clearly have an non-QRT mapping. The question above is answered in Proposition \ref{prop1}  followed by Corollary \ref{cor1}.

In Corollary \ref{cor3}, the same question is answered for the period $4$ map $j$: ${\overline x}=-y,{\overline y}=x,$  where again by composing $j$ with the QRT involution $i_1,$ we have another family of non-QRT maps.

\bpr \label{prop1} For
 \be \label{type_I_hky_integral} I(x,y)=\frac{N(x,y)}{D(x,y)}=\frac{{\bf X}^TA_0{\bf
Y}}{{\bf X}^TA_1{\bf Y}},
\ee
 when $A_1=A_1^T$  and $A_0=-\epsilon A_0^T,$
  there is $I(y,x)+\epsilon I(x,y)=0,$ where $\epsilon= \pm 1.$ \epr
\begin{proof}
When $A_1=A_1^T$ there is:
 $$D(y,x)-D(x,y)={\bf Y}^TA_1{\bf X}-{\bf X}^TA_1{\bf Y}=
{\bf Y}^TA_1{\bf X}-{\bf Y}^TA_1^T{\bf X}={\bf Y}^T(A_1^T-A_1){\bf X}=0.$$
Also for $A_0=-\epsilon A_0^T,$
$$
N(y,x)+\epsilon N(x,y)={\bf Y}^TA_0{\bf X}+\epsilon {\bf X}^TA_0{\bf Y}=
{\bf Y}^TA_0{\bf X}+\epsilon {\bf Y}^TA_0^T{\bf X}={\bf Y}^T(A_0^T+\epsilon A_0){\bf X}=0.
$$
So
 $$\begin{array}{ccl}
 I(y,x)+\epsilon I(x,y)&=&{\ds \frac{N(y,x)}{D(y,x)}+\epsilon \frac{N(x,y)}{D(x,y)}=
\frac{N(y,x) D(x,y)+\epsilon N(x,y)D(y,x)}{D(x,y)D(y,x)}=}\\[4mm]
{}&=&{\ds \frac{D(x,y)\left(N(y,x)+\epsilon N(x,y) \right)}{D(x,y)^2}=0}
\end{array}$$
\epf
\bc \label{cor1} For $A_1=A_1^T$  and $A_0=-\epsilon A_0^T,$ the map
\be \label{type_I_hky} \begin{array}{l}
{\overline x}=y,\\[3mm]
{\overline y}={\ds \frac{f_1(y)-f_2(y)x}{f_2(y)-f_3(y)x}}
\end{array} \quad \mbox{or} \quad
{\overline y}=\frac{f_1(y)-f_2(y){\underline
y}}{f_2(y)-f_3(y){\underline y}},
\ee where
$$
(f_1,f_2,f_3)^T=(A_0{\bf Y})\times (A_1{\bf Y}),
$$
for $\epsilon=-1$ preserves $I(x,y)$ (\ref{type_I_hky_integral}) so is a symmetric QRT   and for $\epsilon=1$ is measure-preserving\footnote{According to \cite{roberts-2}, a mapping of the plane $\phi :$ $(x,y)\rightarrow ({\overline x},{\overline y})$ is (anti) measure-preserving with density $m(x,y),$ if the Jacobian determinant  (J) of $\phi$ can be written as $J=\frac{m(x,y)}{m({\overline x},{\overline y})}$. Anti measure-preservation corresponds to a measure-preserving and orientation-reversing mapping $\phi. $ For all mappings presented in this work, (anti) measure-preservation is proven by direct computation. }  with density $m(x,y)=\left({\bf X}^TA_0{\bf Y}\right)^{-1}$ and preserves $I(x,y)^2$ so is a non-QRT Liouville integrable map \cite{ves1}.
\ec

\br
Non-QRT mappings presented in subsection $(4.2)$ of \cite{quispel-hky} and in section $(2)$ of \cite{gram_ram_11} are subcases of the mapping of the corollary (\ref{cor1}).
\er

\br
QRT mappings with $A_1=A_1^T$  and $A_0=- A_0^T,$ together with their singularity patterns and their connection to discrete Painlev\'e, were studied in \cite{willox1} under the name of antisymmetric QRTs. From the corollary above, it is clear that the even(odd) iterations of the mappings of the corollary (\ref{cor1}), with $\epsilon =1,$ are the antisymmetric QRTs. In other words, we have shown that the \lq\lq square root\rq\rq of the antisymmetric QRTs are non-QRT mappings.
\er

\bex {\em When:
$$
A_0=\left( \begin{array}{ccc}
            (\epsilon-1)\kappa &1 &0\\
            -\epsilon  &(\epsilon-1)\lambda & \gamma\\
            0&-\epsilon \gamma&(\epsilon-1)\mu
            \end{array} \right), \quad  A_1=\left(
            \begin{array}{ccc}
            0&0&0\\
            0&0&0\\
            0&0&1
            \end{array}\right),
$$
according to corollary (\ref{cor1}) there is $$ I(x,y)=x^2 y+\gamma x-\epsilon(x y^2+\gamma y)+(\epsilon-1)(\kappa x^2y^2+\lambda xy+\mu),$$

\begin{eqnarray*}
&&\phi:x_{n+1}=y_n,\quad  y_{n+1}=-x_n+\frac{\epsilon
y_n^2+(\epsilon-1)\lambda y_n-\gamma}{(\epsilon-1)\kappa y_n^2+y_n}, \\
&&\mbox{or}\\
&&x_{n+1}+ x_{n-1}=\frac{\epsilon x_n^2+(\epsilon-1)\lambda x_n-\gamma}{(\epsilon-1)\kappa x_n^2+x_n}.
\end{eqnarray*}
When $\epsilon=-1$ mapping $\phi$ preserves the symmetric integral $I(x,y)$ and is a member of the symmetric QRT.

When $\epsilon=1,$ $I(x,y)$ is antisymmetric under the interchange of its arguments and the mapping $\phi $ preserves $I(x,y)^2$. Hence it is a non-QRT mapping. Moreover the even (odd) iterations of $\phi$ preserves the antisymmetric $I(x,y)$. Hence it is a member of the antisymmetric QRT \cite{willox1}. To see the latter, when $n$ is even, note that $\phi$ reads:
\be \label{oddhky}
x_{2m+1}=y_{2m}, \quad y_{2m+1}=-x_{2m}+y_{2m}-\gamma/y_{2m},
\ee When $n$ is odd, we have:
\be \label{evenhky}
x_{2m}=y_{2m-1}, \quad y_{2m}=-x_{2m-1}+y_{2m-1}-\gamma/y_{2m-1},
\ee
Substituting (\ref{evenhky}) into (\ref{oddhky}) we find that the odd $x$ and $y$ satisfy the antisymmetric QRT map:
$$
x_{2m+1}=-x_{2m-1}+y_{2m-1}-\gamma/y_{2m-1}, \quad y_{2m+1}=-y_{2m-1}+x_{2m+1}-\gamma/x_{2m+1}
$$
Similarly, the even $x$ and $y$ also satisfy the antisymmetric QRT:
$$
x_{2m+2}=-x_{2m}+y_{2m}-\gamma/y_{2m}, \quad y_{2m+2}=-y_{2m}+x_{2m+2}-\gamma/x_{2m+2}.
$$
}\eex

Asking the same questions as above for the period-4 mapping $j$: ${\overline x}=-y,{\overline y}=x,$ we find the following corollary\footnote{The proof  is exactly analogous to the proof of corollary (\ref{cor1}.)}

\bc \label{cor3}
The mapping \be \label{type_I_hky1} \begin{array}{l}
{\overline x}=-y,\\[3mm]
{\overline y}={\ds \frac{f_1(y)-f_2(y)x}{f_2(y)-f_3(y)x}}={\ds \frac{\beta y(\delta y^2+\zeta)+\alpha x(2\delta y^2+\gamma y^4+\zeta)}
{\beta xy(\delta+\gamma y^2)-\alpha(2\delta y^2+\gamma y^4+\zeta)}}
\end{array}\ee or
$$
{\overline y}=\frac{\beta y(\delta y^2+\zeta)-\alpha {\underline y}(2\delta y^2+\gamma y^4+\zeta)}
{-\beta y{\underline y}(\delta+\gamma y^2)-\alpha(2\delta y^2+\gamma y^4+\zeta)},
$$ preserves $I^2$ where $I$ is given by
$$
I(x,y)=\frac{{\bf X}^TA_0{\bf
Y}}{{\bf X}^TA_1{\bf Y}}, \quad \mbox{with} \quad  A_0={\ds\left(\begin{array}{ccc}
                           0&0&\alpha\\
                           0&\beta&0\\
                           -\alpha&0&0
                           \end{array}\right)   } \quad  \mbox{and} \quad  A_1={\ds\left(\begin{array}{ccc}
                           \gamma&0&\delta\\
                           0&0&0\\
                           \delta&0&\zeta
                           \end{array}\right)   },
$$ so is a non-QRT mapping. Mapping (\ref{type_I_hky1}) is anti-measure preserving with density $m(x,y)=\frac{1}{{\bf X}^TA_0{\bf
Y}},$ hence a Liouville integrable map.
\ec

 \section{Asymmetric and Symmetric non-QRT Maps}

Here we present a family of non-QRT mappings that corresponds to asymmetric QRT case, and then we specialise this result to the ones  that correspond to the symmetric QRT case.

\bpr \label{prop04} For
 \be \label{type_I_hky-asym_integral2} I(x,y)=
 \frac{{\bf X}^TA_0{\bf
Y}}{{\bf X}^TA_1{\bf Y}},
 \quad \mbox{where} \quad A_0=\left(\begin{array}{ccc}
\alpha &\beta &\gamma \\
\delta & \epsilon & \zeta \\
\kappa & \lambda & \mu
\end{array}\right), \quad
 A_1=\left(\begin{array}{ccc}
\alpha_1 &\beta_1 &\gamma_1 \\
\delta_1 & \epsilon_1 & \zeta_1 \\
\kappa_1 & \lambda_1 & \mu_1
\end{array} \right),
 \ee with
$$\begin{array}{lll}
\alpha_1=2\alpha^2 f -c \alpha\delta+\beta \delta b-b \alpha \epsilon & \beta_1=2 f \alpha \beta + b \gamma \delta - c \alpha \epsilon -
  b \alpha \zeta & \gamma_1= 2 f \alpha \gamma - c \alpha \zeta\\
\delta_1=2m\alpha^2 - 2 c\alpha\kappa + 2b\beta\kappa -
    2b\alpha\lambda & \epsilon_1=2 m \alpha \beta + 2 b \gamma \kappa - 2 c \alpha \lambda -
  2 b \alpha \mu &\zeta_1=2 m \alpha \gamma - 2 c \alpha \mu\\
  \kappa_1=m \alpha \delta - 2 f \alpha \kappa + b \epsilon \kappa -
  b \delta \lambda & \lambda_1=m \alpha \epsilon + b \zeta \kappa - 2 f \alpha \lambda -
  b \delta \mu &\mu_1=m \alpha \zeta - 2 f \alpha \mu
\end{array}
$$  there is $I\circ i_2=I,$ and $I\circ h_i=-I,$ $i=1,2,$ where $i_2$ the QRT involution and
  $$
h_1:  \left\{\begin{array}{l} {\overline x}=-{\ds\frac{(\delta y^2+\epsilon y+\delta)x+2(\kappa y^2+\lambda y+\mu)}{2(\alpha y^2+\beta y+\gamma)x+\delta y^2+\epsilon y+\delta}},\\
{\overline y}=y,\end{array}\right.$$
$$     h_2:  \left\{\begin{array}{l}
{\overline x}=-{\ds \frac{(\delta_1 y^2+\epsilon_1 y+\delta_1)x+2(\kappa_1 y^2+\lambda_1 y+\mu_1)}{2(\alpha_1 y^2+\beta_1 y+\gamma_1)x+\delta_1 y^2+\epsilon_1 y+\delta_1}},\\
{\overline y}=y,
\end{array} \right.
    $$ the non-QRT involutions.
   \epr
\bpf
Lets consider the function:
\be \label{mas1}
I(x,y,{\overline x})=\frac{x{\overline x}(\alpha y^2+\beta y+\gamma)+x(\rho y^2+\sigma y +\tau)+{\overline x}(\delta y^2+\epsilon y+\zeta)+\kappa y^2+\lambda y+\mu}
{ x{\overline x}(a y^2 +b y+c)+(x+{\overline x})(d y^2+e y +f)+k y^2+l y+m},
\ee
where $\alpha,\beta,\ldots , l,m $ are parameters and ${\overline x}=f(x,y)$ is a  function  of $x,y,$ to be determined to be an involution, i.e., ${\overline {\overline x}}=x$. Now the solution of the equation $I({\overline x},y,x)+I(x,y,{\overline x})=0,$ for ${\overline x}$ will determine the form of the latter. Explicitly ${\overline x}$ reads:
 $$
 {\overline x}=-\frac{x\left(y^2(\delta+\rho)+y(\epsilon+\sigma)+\zeta+\tau\right)+2(\kappa y^2+\lambda y+\mu)}
 {2x (\alpha y^2+\beta y+\gamma)+y^2(\delta+\rho)+y(\epsilon+\sigma)+\zeta+\tau}.
 $$
 Substituting this solution back to $(\ref{mas1}),$ we get $I=I(x,y)$ which is quadratic in $x$ and of higher degree than quadratic in $y.$ Clearly, the equation $I({\overline x},y)+I(x,y)=0$ by construction has as a solution the previously defined $({\tilde x})$. A choice of the parameters of $(\ref{mas1}),$ that will lead to biquadratic $I(x,y)$ is $\rho=\delta,\sigma=\epsilon,\tau=0,a=d=k=0,e=\frac{b\delta}{\alpha},l=\frac{b\kappa}{\alpha}.$ Then and after the scaling $(\delta,\epsilon)\rightarrow (\delta/2,\epsilon/2),$ the integral ${\ds I(x,y)=\frac{{\bf X}^TA_0{\bf
Y}}{{\bf X}^TA_1{\bf Y}}}$ gets the form above and the solution of $I({\overline x},y)+I(x,y)=0$ is exactly $h_1,h_2.$

\epf

\bc \label{cor4a}
The mapping \be \label{type_I_hky1_asym} \begin{array}{l}
{\overline x}=-{\ds\frac{(\delta y^2+\epsilon y+\delta)x+2(\kappa y^2+\lambda y+\mu)}{2(\alpha y^2+\beta y+\gamma)x+\delta y^2+\epsilon y+\delta}},\\[3mm]
{\overline y}={\ds \frac{g_1({\overline x})-y g_2({\overline x})}{g_2({\overline x})-yg_3({\overline x})}}
\end{array}\ee
and the dual mapping
\be \label{type_I_hky1_asym_dual} \begin{array}{l}
{\overline x}=-{\ds \frac{(\delta_1 y^2+\epsilon_1 y+\delta_1)x+2(\kappa_1 y^2+\lambda_1 y+\mu_1)}{2(\alpha_1 y^2+\beta_1 y+\gamma_1)x+\delta_1 y^2+\epsilon_1 y+\delta_1}},\\[3mm]
{\overline y}={\ds \frac{g_1({\overline x})-y g_2({\overline x})}{g_2({\overline x})-yg_3({\overline x})}}
\end{array}\ee
preserve $I^2$ where $I$ is given by (\ref{type_I_hky-asym_integral2}), so they are non-QRT maps. Mapping (\ref{type_I_hky1_asym}) is measure-preserving, whereas mapping (\ref{type_I_hky1_asym_dual}) is anti measure-preserving, both with density $m(x,y)=\frac{1}{{\bf X}^TA_0{\bf Y}},$ hence Liouville integrable maps.

\ec

If we now demand  the matrices of $(\ref{type_I_hky-asym_integral2})$ to be symmetric, we have the following corollary:

%\bpr \label{prop4} For
% \be \label{type_I_hky-sym_integral2} I(x,y)=\frac{N(x,y)}{D(x,y)}=
% \frac{{\bf X}^TA_0{\bf
%Y}}{{\bf X}^TA_1{\bf Y}},
%\ee with
%$$
%A_0=\left(\begin{array}{ccc}
%\alpha &\beta &\gamma \\
%\beta & \epsilon & \delta \\
%\gamma & \delta & \mu
%\end{array}\right), \quad
% A_1=\left(\begin{array}{ccc}
%\alpha_1 &\beta_1 &\gamma_1 \\
%\beta_1 & \epsilon_1 & \delta_1 \\
%\gamma_1 & \delta_1 & \mu_1
%\end{array} \right),
%$$
%where
%$$\begin{array}{ccc}
%\alpha_1=\alpha \epsilon -\beta^2 & \beta_1=2 \alpha \delta -2\beta \gamma & \gamma_1= \beta \delta -\gamma \epsilon\\
%\epsilon_1=4\alpha \mu-4 \gamma^2 & \delta_1=2\beta \mu -2\gamma
%\delta & \mu_1=\epsilon \mu-\delta^2
%\end{array}$$
 % there is $I\circ j=I,$ and $I\circ h_i=-I,$ where $j: {\overline x}=y,$ ${\overline y}=x,$ and
 % $$
%h_1:  \left\{\begin{array}{l} {\ds {\overline x}= -\frac{2\gamma y^2+2\delta y+2\mu
%+x(\beta y^2+\epsilon y+\delta)}{\beta y^2+\epsilon
%y+\delta+2x(\alpha y^2+\beta y+\gamma)}}\\[3mm]
% {\overline y}=y, \end{array} \right.
%     h_2:  \left\{\begin{array}{l} {\ds {\overline x}= -\frac{2\gamma_1 y^2+2\delta_1 y+2\mu_1
%+x(\beta_1 y^2+\epsilon_1 y+\delta_1)}{\beta_1 y^2+\epsilon_1
%y+\delta_1+2x(\alpha_1 y^2+\beta_1 y+\gamma_1)}}\\[3mm]
% {\overline y}=y, \end{array} \right.
%    $$
 %  \epr
%\bpf
%The validity of $I\circ j=I,$ follows from the fact that the matrices $A_0,$  $A_1$  are symmetric hence  $I(x,y)$ is preserved by the involution %$j$.
%The validity of $I\circ h_i=-I,$  can be proven by direct computation.
%\epf

\bc \label{cor4}
The mapping \be \label{type_I_hky1_sym} \begin{array}{l}
{\overline x}=y,\\[3mm]
{\overline y}={\ds -\frac{2\gamma y^2+2\delta y+2\mu
+x(\beta y^2+\epsilon y+\delta)}{\beta y^2+\epsilon
y+\delta+2x(\alpha y^2+\beta y+\gamma)}}
\end{array}\ee or
$$
{\overline y}=-\frac{2\gamma y^2+2\delta y+2\mu
+{\underline y}(\beta y^2+\epsilon y+\delta)}{\beta y^2+\epsilon
y+\delta+2{\underline y}(\alpha y^2+\beta y+\gamma)},$$
and the dual mapping
\be \label{type_I_hky1_sym_dual} \begin{array}{l}
{\overline x}=y,\\[3mm]
{\overline y}={\ds -\frac{2\gamma_1 y^2+2\delta_1 y+2\mu_1
+x(\beta_1 y^2+\epsilon_1 y+\delta_1)}{\beta_1 y^2+\epsilon_1
y+\delta_1+2x(\alpha_1 y^2+\beta_1 y+\gamma_1)}}
\end{array}\ee or
$$
{\overline y}=-\frac{2\gamma_1 y^2+2\delta_1 y+2\mu_1
+{\underline y}(\beta_1 y^2+\epsilon_1 y+\delta_1)}{\beta_1 y^2+\epsilon_1
y+\delta_1+2{\underline y}(\alpha_1 y^2+\beta_1 y+\gamma_1)},$$
preserve $I^2$ where $I$ is given by
\be \label{type_I_hky-sym_integral2}
I(x,y)=
 \frac{{\bf X}^TA_0{\bf
Y}}{{\bf X}^TA_1{\bf Y}},
\quad \mbox{ where} \quad
A_0=\left(\begin{array}{ccc}
\alpha &\beta &\gamma \\
\beta & \epsilon & \delta \\
\gamma & \delta & \mu
\end{array}\right), \quad
 A_1=\left(\begin{array}{ccc}
\alpha_1 &\beta_1 &\gamma_1 \\
\beta_1 & \epsilon_1 & \delta_1 \\
\gamma_1 & \delta_1 & \mu_1
\end{array} \right),
\ee
with
$$\begin{array}{lll}
\alpha_1=\alpha \epsilon -\beta^2 & \beta_1=2 \alpha \delta -2\beta \gamma & \gamma_1= \beta \delta -\gamma \epsilon\\
\epsilon_1=4\alpha \mu-4 \gamma^2 & \delta_1=2\beta \mu -2\gamma
\delta & \mu_1=\epsilon \mu-\delta^2
\end{array}$$
 so they are non-QRT maps. Mapping (\ref{type_I_hky1_sym}) is anti measure-preserving, whereas mapping (\ref{type_I_hky1_sym_dual}) is  measure-preserving, both with density $m(x,y)=\frac{1}{{\bf X}^TA_0{\bf Y}},$ hence Liouville integrable maps.
\ec

 \br For $\alpha=0,$ $\beta=0,$
$\gamma=1/2,$ $\delta=0,$ $\epsilon=a,$ $\mu=b/2,$  the integral (\ref{type_I_hky-sym_integral2})
takes  the form:
$$
I({\underline y},y)=\frac{2a{\underline y}y+{\underline y}^2+y^2+b}{2{\underline y}y+a({\underline y}^2+y^2)-ab}
$$ that the associate mapping (\ref{type_I_hky1_sym}) is  the non-QRT mapping
associated with $Q3$ found in \cite{Nalini1hky}.
 \er

 \br For $\alpha=k^2sn(a)sn(b)sn(a-b),$ $\beta=0,$ $\gamma=-sn(a-b),$ $\delta=0,$
$\epsilon=2 (sn(a)-sn(b)),$ $\mu=sn(a)sn(b)sn(a-b),$  the integral (\ref{type_I_hky-sym_integral2})
takes the form:
\begin{eqnarray*}
&&I({\underline y},y)=\\
&&\frac{((1+k^2{\underline y}^2y^2)sn(a)sn(b)-{\underline y}^2-y^2)sn(a-b)+2{\underline y}y(sn(a)-sn(b))}
{((1+k^2{\underline y}^2y^2)sn(a)sn(b)+{\underline y}^2+y^2)(sn(a)-sn(b))+2{\underline y}ysn(a-b)(k^2sn(a)^2sn(b)^2-1)}
\end{eqnarray*}
that the associate mapping (\ref{type_I_hky1_sym}) is  the non-QRT mapping
associated with $Q4$ found in \cite{Nalini1hky}.
 \er
\bex
{\em
For $\alpha=1,$ $\beta=\delta=0,$ $I$  given by (\ref{type_I_hky-sym_integral2}) takes the form:
$$
I(x,y)=\frac{x^2y^2+\gamma (x^2+y^2)+\mu}{xy}, \quad \mbox{or} \quad I({\underline y},y)=\frac{{\underline y}^2y^2+\gamma ({\underline y}^2+y^2)+\mu}{{\underline y}y},
$$ and from (\ref{type_I_hky1_sym}) there is:
\be \label{hky-qpIII}
{\overline y}{\underline y}=-\frac{\mu+\gamma y^2}{y^2+\gamma}
\ee that preserves $I({\underline y},y)^2.$ Inspired by \cite{hone1}, a discrete Lax pair for (\ref{hky-qpIII}) is given by
\be \label{lax-pair}
{\bf L}_n\Psi_n=\lambda \Psi_n, \quad \Psi_{n+1}={\bf M}_n\Psi_n,
\ee
where
$$
{\bf L}_n(\lambda)=\left(\begin{array}{ll}
(-1)^ny_{n-1}y_n &-\gamma\\
\lambda-y_{n-1}^2-y_n^2&(-1)^n\frac{\mu+\gamma (y_{n-1}^2+y_n^2)}{y_{n-1}y_n}
\end{array}\right),$$
$$  {\bf M}_n(\lambda)=\left(\begin{array}{ll}
0&-\gamma \\
\lambda-y_{n-1}^2-y_n^2&(-1)^n\gamma\frac{\mu+\gamma y_n^2+y_{n-1}^2(\gamma +y_n^2)}{y_{n-1}y_n(\gamma +y_n^2)}
\end{array}\right).
$$
The equation (\ref{hky-qpIII}) arises as the compatibility condition ${\bf L}_{n+1}{\bf M}_n-{\bf M}_n{\bf L}_n=0$ for the linear system (\ref{lax-pair}). }
\eex

\subsection{Integration of  the non-QRT mappings}

 In exactly the same manner as in \cite{gram-int-QRT}, and \cite{iatrou2}, by integration of a non-QRT mapping we will mean the parametrisation of the corresponding invariant curves.

 In this setting, the integration of the symmetric QRT,  according to Veselov \cite{ves1} was first considered by Euler, but of course not inside the modern context of integrability. A more transparent and explicit method was presented by Baxter \cite{baxt1}. Recently, the authors of \cite{gram-int-QRT}, \cite{iatrou2}, independently presented the integration of the asymetric QRT.

To proceed with the integration of the non-QRT mappings (\ref{type_I_hky}) and (\ref{type_I_hky1}), we should first make the following observation.
 Let us consider $${\ds I(x_{n-1},x_n;n)=(-1)^n\frac{{\bf X}_{n-1}^TA_0{\bf
X}_n}{{\bf X}_{n-1}^TA_1{\bf X}_n}},$$ then
$$
I(x_{n},x_{n+1};n+1)-I(x_{n-1},x_n;n)=(-1)^{n+1}\left(\frac{{\bf X}_{n}^TA_0{\bf
X}_{n+1}}{{\bf X}_{n}^TA_1{\bf X}_{n+1}}+\frac{{\bf X}_{n-1}^TA_0{\bf
X}_n}{{\bf X}_{n-1}^TA_1{\bf X}_n} \right),
$$ that is satisfied by mappings (\ref{type_I_hky}) and (\ref{type_I_hky1}). Then the invariant curves that the HKY under consideration lies is:
\be \label{incurves}
{\bf X}_{n-1}^TA_0{\bf
X}_n-(-1)^n K {\bf X}_{n-1}^TA_1{\bf X}_n=0,
\ee where $K$ is the integration constant associated with the initial conditions $(x_{-1},x_0),$ $K=I(x_{-1},x_0)$. Then, after defining
$M=A_0-(-1)^n KA_1,$ the way to parametrise ${\bf X}_{n-1}^TM{\bf X}_n=0,$  is exactly the one presented in \cite{gram-int-QRT} for the asymmetric QRT.

\section{Conclusions}
In the previous sections, multiparameter mappings of non-QRT type were presented. By construction they preserve a bi-quartic invariant that can be considered as the square of a QRT bi-quadratic one.  Apart from the preservation of a bi-quartic integral, these systems turn to be (anti) measure-preserving and hence Liouville integrable.

Commenting on the search for general solution of such integrable systems, it should be noted  that by fixing an initial condition say $(x_{-1},x_0),$ we have $K=I(x_{-1},x_0)$ and equation (\ref{incurves}) defines, in general, two elliptic curves: one for $n$ odd and the other for $n$ even. The non-QRT system at each step maps one curve to the other and vice versa and since the latter is bi-rational, these two curves are birationally equivalent  and hence they should have the same j-invariant.  In this sense we should not expect to find the general solution of a non-QRT map as an addition formula on one elliptic curve since the latter simply alternates between two birationally equivalent elliptic curves. Also in this setting by construction, by finding a non-QRT map we have two birationally equivalent curves and the non-QRT map is the birational relation between these two curves. Note also that these are  special classes of  birationally equivalent curves since the birational relation that maps one to the other and the inverse of it are  the same birational map and the one we call the non-QRT map.

It should be pointed out, that although general multiparameter families of these systems are presented,  we have by no means exhausted all possble non-QRT mappings of the plane. A natural extension of this work is to consider systems whose higher iterates are QRT maps. If the third iteration is a QRT mapping, then they preserve $I^3,$ with $I$ associated to the QRT bi-quadratic invariant. This will be considered elsewhere.

\subsection*{Acknowledgements}

The research reported here was supported by the Australian Research Council  Discovery Project Grant \#DP0664624. The authors would like to thank the Isaac Newton Institute for Mathematical Sciences, where this paper was completed, for its hospitality.

\bibliographystyle{alpha}

\end{document}